\documentclass[aps,prl,superscriptaddress,showpacs,floatfix,twocolumn]{revtex4}
\usepackage{graphicx}

\newcommand{\mean}[1]{\left\langle #1 \right\rangle} 

\begin{document}


\title{Elliptic flow for $\phi$ mesons and (anti)deuterons 
    in Au+Au collisions at $\sqrt{s_{NN}}$ = 200 GeV
}

\newcommand{\abilene}{Abilene Christian University, Abilene, TX 79699, U.S.}
\newcommand{\banaras}{Department of Physics, Banaras Hindu University, Varanasi 221005, India}
\newcommand{\bnl}{Brookhaven National Laboratory, Upton, NY 11973-5000, U.S.}
\newcommand{\caucr}{University of California - Riverside, Riverside, CA 92521, U.S.}
\newcommand{\cns}{Center for Nuclear Study, Graduate School of Science, University of Tokyo, 7-3-1 Hongo, Bunkyo, Tokyo 113-0033, Japan}
\newcommand{\colorado}{University of Colorado, Boulder, CO 80309, U.S.}
\newcommand{\columbia}{Columbia University, New York, NY 10027 and Nevis Laboratories, Irvington, NY 10533, U.S.}
\newcommand{\dapnia}{Dapnia, CEA Saclay, F-91191, Gif-sur-Yvette, France}
\newcommand{\debrecen}{Debrecen University, H-4010 Debrecen, Egyetem t{\'e}r 1, Hungary}
\newcommand{\elte}{ELTE, E{\"o}tv{\"o}s Lor{\'a}nd University, H - 1117 Budapest, P{\'a}zm{\'a}ny P. s. 1/A, Hungary}
\newcommand{\fsu}{Florida State University, Tallahassee, FL 32306, U.S.}
\newcommand{\gsu}{Georgia State University, Atlanta, GA 30303, U.S.}
\newcommand{\hiroshima}{Hiroshima University, Kagamiyama, Higashi-Hiroshima 739-8526, Japan}
\newcommand{\ihepprot}{IHEP Protvino, State Research Center of Russian Federation, Institute for High Energy Physics, Protvino, 142281, Russia}
\newcommand{\illuiuc}{University of Illinois at Urbana-Champaign, Urbana, IL 61801, U.S.}
\newcommand{\isu}{Iowa State University, Ames, IA 50011, U.S.}
\newcommand{\jinrdubna}{Joint Institute for Nuclear Research, 141980 Dubna, Moscow Region, Russia}
\newcommand{\kaeri}{KAERI, Cyclotron Application Laboratory, Seoul, South Korea}
\newcommand{\kek}{KEK, High Energy Accelerator Research Organization, Tsukuba, Ibaraki 305-0801, Japan}
\newcommand{\kfki}{KFKI Research Institute for Particle and Nuclear Physics of the Hungarian Academy of Sciences (MTA KFKI RMKI), H-1525 Budapest 114, POBox 49, Budapest, Hungary}
\newcommand{\korea}{Korea University, Seoul, 136-701, Korea}
\newcommand{\kurchatov}{Russian Research Center ``Kurchatov Institute", Moscow, Russia}
\newcommand{\kyoto}{Kyoto University, Kyoto 606-8502, Japan}
\newcommand{\labllr}{Laboratoire Leprince-Ringuet, Ecole Polytechnique, CNRS-IN2P3, Route de Saclay, F-91128, Palaiseau, France}
\newcommand{\lawllnl}{Lawrence Livermore National Laboratory, Livermore, CA 94550, U.S.}
\newcommand{\losalamos}{Los Alamos National Laboratory, Los Alamos, NM 87545, U.S.}
\newcommand{\lpc}{LPC, Universit{\'e} Blaise Pascal, CNRS-IN2P3, Clermont-Fd, 63177 Aubiere Cedex, France}
\newcommand{\lund}{Department of Physics, Lund University, Box 118, SE-221 00 Lund, Sweden}
\newcommand{\muenster}{Institut f\"ur Kernphysik, University of Muenster, D-48149 Muenster, Germany}
\newcommand{\myongji}{Myongji University, Yongin, Kyonggido 449-728, Korea}
\newcommand{\nagasaki}{Nagasaki Institute of Applied Science, Nagasaki-shi, Nagasaki 851-0193, Japan}
\newcommand{\newmex}{University of New Mexico, Albuquerque, NM 87131, U.S. }
\newcommand{\nmsu}{New Mexico State University, Las Cruces, NM 88003, U.S.}
\newcommand{\ornl}{Oak Ridge National Laboratory, Oak Ridge, TN 37831, U.S.}
\newcommand{\orsay}{IPN-Orsay, Universite Paris Sud, CNRS-IN2P3, BP1, F-91406, Orsay, France}
\newcommand{\pnpi}{PNPI, Petersburg Nuclear Physics Institute, Gatchina, Leningrad region, 188300, Russia}
\newcommand{\riken}{RIKEN, The Institute of Physical and Chemical Research, Wako, Saitama 351-0198, Japan}
\newcommand{\rikjrbrc}{RIKEN BNL Research Center, Brookhaven National Laboratory, Upton, NY 11973-5000, U.S.}
\newcommand{\rikkyo}{Physics Department, Rikkyo University, 3-34-1 Nishi-Ikebukuro, Toshima, Tokyo 171-8501, Japan}
\newcommand{\saispbstu}{Saint Petersburg State Polytechnic University, St. Petersburg, Russia}
\newcommand{\saopaulo}{Universidade de S{\~a}o Paulo, Instituto de F\'{\i}sica, Caixa Postal 66318, S{\~a}o Paulo CEP05315-970, Brazil}
\newcommand{\seoulnat}{System Electronics Laboratory, Seoul National University, Seoul, South Korea}
\newcommand{\stonybrkc}{Chemistry Department, Stony Brook University, Stony Brook, SUNY, NY 11794-3400, U.S.}
\newcommand{\stonycrkp}{Department of Physics and Astronomy, Stony Brook University, SUNY, Stony Brook, NY 11794, U.S.}
\newcommand{\subatech}{SUBATECH (Ecole des Mines de Nantes, CNRS-IN2P3, Universit{\'e} de Nantes) BP 20722 - 44307, Nantes, France}
\newcommand{\tenn}{University of Tennessee, Knoxville, TN 37996, U.S.}
\newcommand{\titech}{Department of Physics, Tokyo Institute of Technology, Oh-okayama, Meguro, Tokyo 152-8551, Japan}
\newcommand{\tsukuba}{Institute of Physics, University of Tsukuba, Tsukuba, Ibaraki 305, Japan}
\newcommand{\vandy}{Vanderbilt University, Nashville, TN 37235, U.S.}
\newcommand{\waseda}{Waseda University, Advanced Research Institute for Science and Engineering, 17 Kikui-cho, Shinjuku-ku, Tokyo 162-0044, Japan}
\newcommand{\weizmann}{Weizmann Institute, Rehovot 76100, Israel}
\newcommand{\yonsei}{Yonsei University, IPAP, Seoul 120-749, Korea}
\affiliation{\abilene}
\affiliation{\banaras}
\affiliation{\bnl}
\affiliation{\caucr}
\affiliation{\cns}
\affiliation{\colorado}
\affiliation{\columbia}
\affiliation{\dapnia}
\affiliation{\debrecen}
\affiliation{\elte}
\affiliation{\fsu}
\affiliation{\gsu}
\affiliation{\hiroshima}
\affiliation{\ihepprot}
\affiliation{\illuiuc}
\affiliation{\isu}
\affiliation{\jinrdubna}
\affiliation{\kaeri}
\affiliation{\kek}
\affiliation{\kfki}
\affiliation{\korea}
\affiliation{\kurchatov}
\affiliation{\kyoto}
\affiliation{\labllr}
\affiliation{\lawllnl}
\affiliation{\losalamos}
\affiliation{\lpc}
\affiliation{\lund}
\affiliation{\muenster}
\affiliation{\myongji}
\affiliation{\nagasaki}
\affiliation{\newmex}
\affiliation{\nmsu}
\affiliation{\ornl}
\affiliation{\orsay}
\affiliation{\pnpi}
\affiliation{\riken}
\affiliation{\rikjrbrc}
\affiliation{\rikkyo}
\affiliation{\saispbstu}
\affiliation{\saopaulo}
\affiliation{\seoulnat}
\affiliation{\stonybrkc}
\affiliation{\stonycrkp}
\affiliation{\subatech}
\affiliation{\tenn}
\affiliation{\titech}
\affiliation{\tsukuba}
\affiliation{\vandy}
\affiliation{\waseda}
\affiliation{\weizmann}
\affiliation{\yonsei}
\author{S.~Afanasiev}	\affiliation{\jinrdubna}
\author{C.~Aidala}	\affiliation{\columbia}
\author{N.N.~Ajitanand}	\affiliation{\stonybrkc}
\author{Y.~Akiba}	\affiliation{\riken} \affiliation{\rikjrbrc}
\author{J.~Alexander}	\affiliation{\stonybrkc}
\author{A.~Al-Jamel}	\affiliation{\nmsu}
\author{K.~Aoki}	\affiliation{\kyoto} \affiliation{\riken}
\author{L.~Aphecetche}	\affiliation{\subatech}
\author{R.~Armendariz}	\affiliation{\nmsu}
\author{S.H.~Aronson}	\affiliation{\bnl}
% ...
\collaboration{PHENIX Collaboration} \noaffiliation

\date{\today}


\begin{abstract}

Differential elliptic flow ($v_2$) for $\phi$ mesons and 
(anti)deuterons $(\overline{d})d$ is measured for Au+Au collisions 
at $\sqrt{s_{NN}}~=~200$ GeV. The $v_2$ for $\phi$ mesons follows 
the trend of lighter $\pi^{\pm}$ and $K^{\pm}$ mesons, suggesting 
that ordinary hadrons interacting with standard hadronic cross 
sections are not the primary driver for elliptic flow development. 
The $v_2$ values for $(\overline{d})d$ suggest that elliptic flow is 
additive for composite particles.  This further validation of the 
universal scaling of $v_2$ per constituent quark for baryons and 
mesons suggests that partonic collectivity dominates the transverse 
expansion dynamics.

\end{abstract}


\maketitle


An important goal of current ultra-relativistic heavy ion research is to 
map out the accessible regions of the Quantum Chromodynamics (QCD) phase 
diagram. Central to this goal, is the creation and study of a new phase of 
nuclear matter -- the Quark Gluon Plasma (QGP). Thermalization and 
de-confinement are important properties of this matter, believed to be 
produced in heavy ion collisions at the Relativistic Heavy Ion Collider 
(RHIC)~\cite{Adcox:2004mh,Gyulassy:2004zy,Shuryak:2004cy}.
	
Detailed elliptic flow measurements provide indispensable information 
about this high energy density 
matter~\cite{Ackermann:2000tr,Adcox:2002ms,Adler:2003kt,Adler:2004cj,Lacey:2005qq}. 
Such measurements are characterized by the magnitude of the 
second-harmonic coefficient
$v_2 = \mean{e^{i2(\varphi_p - \Phi_{RP})}}$,  
of the Fourier expansion of the azimuthal distribution of emitted 
particles. Here, $\varphi_{p}$ represents the azimuthal emission angle of 
a particle, $\Phi_{RP}$ is the azimuthal angle of the reaction plane and 
the brackets denote statistical averaging over particles and 
events~\cite{Demoulins90,Voloshin96}.
	
At RHIC energies, there is now significant evidence that elliptic flow, in 
non-central collisions, results from hydrodynamic pressure gradients 
developed in a locally thermalized ``almond-shaped" collision zone. That 
is, the initial transverse coordinate-space anisotropy of this zone is 
converted, via particle interactions, into an azimuthal momentum-space 
anisotropy. Indeed, when plotted as a function of the transverse kinetic 
energy $KE_T \equiv m_T-m$ divided by the number of valence quarks $n_q$, 
of a given hadron ($n_q = 2$ for mesons and $n_q = 3$ for baryons), 
$v_2/n_q$ shows universal scaling for a broad range of particle 
species~\cite{Issah:2006qn,Adare:2006ti,Lacey:2006pn} 
($m_T$ is the transverse mass). This has been interpreted as evidence that 
hydrodynamic expansion of the QGP occurs during a phase characterized by 
(i) a rather low viscosity to entropy ratio 
$\eta/s$~\cite{Shuryak:2004cy,Gyulassy:2004zy,Heinz:2001xi,Asakawa:2006tc,Lacey:2006pn} 
and (ii) independent quasi-particles which exhibit the quantum numbers of 
quarks~\cite{Voloshin:2002wa,Fries:2003kq,Greco:2003mm,Xu:2005jt,Muller:2006ee,Lacey:2006pn}. 
A consensus on the detailed dynamical evolution of the QGP has not been 
reached~\cite{Shuryak:2004cy,Asakawa:2006tc}.

Elliptic flow measurements for heavy, strange and multi-strange 
hadrons~\cite{Oldenburg:2006br,Adare:2006nq} can lend unique insight on 
reaction dynamics. Here, we use differential $v_2$ measurements for the 
$\phi$ meson and the deuteron 
to address the important question of how the 
existence of a hadronic phase affects $v_2$, i.e whether or not elliptic 
flow development is dominantly pre- or post-hadronization.
		
The $\phi$ meson is comprised of a strange ($s$) and an anti-strange 
($\bar{s}$) quark and its interaction with hadrons is suppressed according 
to the Okubo-Zweig-Izuka (OZI) rules~\cite{Okubo:1963fa}. One consequence 
of this is that the $\phi$ meson is expected to have a rather small 
hadronic cross section with non-strange hadrons 
($\sim 9$~mb)~\cite{Shor:1984ui,Ko:1993id,Haglin:1994xu}. 
Such a cross-section leads to a relatively large mean free path 
$\lambda_\phi$, when compared to the transverse size of the emitting 
system~\cite{Muller:2006ee,Lacey:2006pn}. Thus, if elliptic flow was 
established in a phase involving hadrons interacting with their standard 
hadronic cross sections (post-hadronization), one would expect $v_2$ for 
the $\phi$ meson to be significantly smaller than that for other hadrons 
(e.g. $p$ and $\pi$).
If $v_2$ is established in the phase prior to hadronization, the $\phi$ 
meson provides an important benchmark test for universal scaling in that 
its mass is similar to that of the proton and the $\Lambda$ baryon, but 
its $v_2$ should be additive with respect to the $v_2$ of its two 
constituent quarks (i.e. $n_q = 2$). Therefore, a detailed comparison of 
the $v_2$ values for the $\phi$ meson with those for other particle 
species, comprised of the lighter $u$ and $d$ quarks or the heavier charm 
quark $c$, can provide unique insight on whether or not partonic 
collectivity plays a central role in reaction dynamics at 
RHIC~\cite{Nonaka:2003ew,Xu:2005jt,Chen:2005nw}.

The deuteron is a very shallow composite $p+n$ bound state, whose binding 
energy ($\sim$ 2.24 MeV) is much less than the hadronization temperature. 
Thus, it is likely that it would suffer from medium induced breakup in the 
hadronic phase, even if it was produced at hadronization. In fact, recent 
investigations~\cite{Adler:2001uy,Adler:2004uy} 
suggest that $(\overline{p} \overline{n})pn$ coalescence dominates the 
(anti)deuteron $(\overline{d})d$ yield in Au+Au collisions. Thus, $v_2$ 
measurements for $(\overline{d})d$ also provide an important test for the 
universal scaling of elliptic flow~\cite{Nonaka:2003ew} in that its $v_2$ 
should be additive; first, with respect to the $v_2$ of its constituent 
hadrons and second, with respect to the $v_2$ of the constituent quarks of 
these hadrons, i.e. $n_q = 2\times 3$.

In the 2004 running period the PHENIX detector~\cite{Adcox:2003zm} 
recorded $\approx 6.5 \times 10^8$ minimum-bias events 
for Au+Au collisions at $\sqrt{s_{NN}}=$ 200~GeV.
The collision vertex $z$ (along beam axis) 
was constrained to 
$|z|~<$~30~cm of the 
nominal crossing point.  The event centrality was determined via cuts in 
the space of Beam-Beam Counter (BBC) charge versus Zero Degree Calorimeter 
energy~\cite{Adcox:2003nr}.
In the central rapidity region ($|\eta|\leq 0.35$) 
the drift chambers, 
each with an azimuthal coverage 
$\Delta\varphi=\pi/2$, and two layers of multi-wire proportional chambers 
with pad readout (PC1 and PC3) were used for charged particle tracking and 
momentum reconstruction.  
The time-of-flight (TOF) and lead scintillator 
(PbSc) detectors were used for charged particle 
identification~\cite{Adler:2003kt,Adler:2004cj}.

Time-of-flight measurements from the TOF and PbSc were used 
in conjunction with the measured momentum and flight-path length, to 
generate a mass-squared ($m^2$) distribution~\cite{Adler:2003cb}.
A track confirmation hit within a $2.5 
\sigma$ matching window in PC3 or TOF/PbSc served to eliminate most 
albedo, conversions, and resonance decays.
A momentum dependent $\pm 2\sigma$ cut about each peak in the $m^2$ 
distribution was used to identify pions ($\pi^{\pm}$), kaons ($K^{\pm}$), 
(anti)protons ($(\overline{p})p$), and ($(\overline{d})d$) in the range 
$0.2<p_T<2.5$ GeV/$c$, $0.3<p_T<2.5$ GeV/$c$, $0.5<p_T<4.5$ GeV/$c$, and 
$1.1<p_T<4.5$ GeV/$c$ respectively in the TOF, and to identify $K^{\pm}$ 
in the range $0.3<p_T<1.5$ GeV/$c$ in the PbSc. This gives $\sim 59000$ 
$\overline{d}+d$. An invariant mass analysis of the $\phi\rightarrow 
K^{+}K^{-}$ decay channel yielded $\sim 340000$ $\phi$ mesons with 
relatively good signal to background (14 - 42\% for the mass window 
$\left| m_{inv}\right|$ = 5 MeV/$c^2$ about the $\phi$ meson peak) over 
the range $1.0 <p_T< 5.5$ GeV/$c$ for $K^{+}K^{-}$ pairs.

The reaction plane method~\cite{Adler:2003kt}
was used to correlate 
the azimuthal angles of charged tracks
with the azimuth of the event plane $\Phi_{2}$, determined via 
hits in the two BBCs 
covering the pseudo-rapidity range $3 < \left| \eta \right| < 3.9 $. The 
large rapidity gap $\Delta\eta > 2.75$ between the central arms and the 
particles used for reaction plane determination reduces the influence of 
possible non-flow contributions, especially those from 
di-jets~\cite{Jia:2006sb}.

\begin{figure}[tb]
\includegraphics[width=1.0\linewidth]{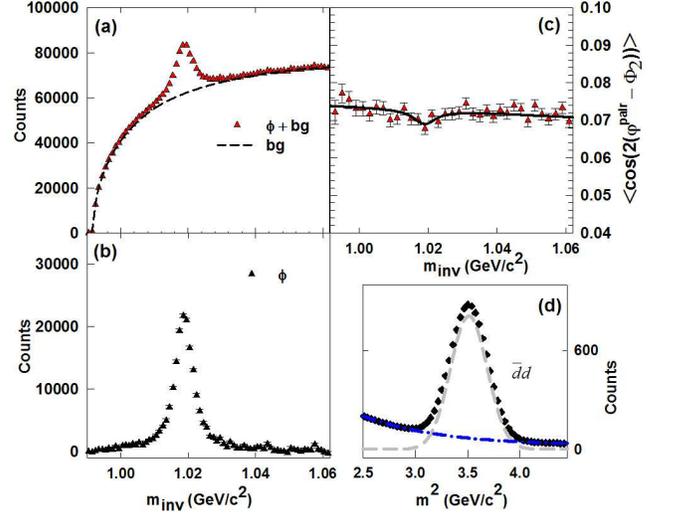}
\caption{(a) $m_{inv}$ distributions for foreground (points) and 
background (dashed-line) $K^{+}K^{-}$ pairs ($p_{T}^{pair} = 1.6-2.7$ GeV/$c$)
for 20-60\% central Au+Au collisions.
(b) $m_{inv}$ distribution after subtraction of the background; 
(c) $\left<cos\left(2(\varphi^{pair}-\Phi_{2})\right)\right>$ vs. $m_{inv}$;
the solid line is a fit to the data with Eq.~\ref{v2fit}.
(d) $m^2$ distribution for $\overline{d},d$ for $p_T = 1.6-2.9$ GeV/$c$.
}
\label{Fig1_inv_mass_v2}
\end{figure}

Charge averaged values of $v_2 = {\left\langle \cos(2(\varphi_p 
-\Phi_2))\right\rangle}/ 
{\left\langle\cos(2(\Phi_2-\Phi_{RP}))\right\rangle}$ were evaluated for 
$\pi^{\pm}$, $K^{\pm}$, $(\overline{p})p$, and $(\overline{d})d$. Here, 
the denominator represents a resolution factor which corrects for the 
difference between the estimated $\Phi_{2}$ and the true azimuth 
$\Phi_{RP}$ of the reaction plane \cite{Poskanzer:1998yz,Adler:2003kt}. 
The estimated resolution factor of the combined reaction plane from both 
BBC's has an average of 0.33 over centrality, with a maximum of about 0.42 
in mid-central collisions~\cite{Adler:2003kt,Adare:2006ti}. The associated 
systematic error is estimated to be $\sim 5$\% for $\pi^{\pm}$, $K^{\pm}$, 
and $(\overline{p})p$. A $p_T$ dependent correction factor ($\sim 5 - 
11\%$) was applied to the $v_2$ values for $(\overline{d})d$, to account 
for background contributions to the (anti)deuteron peak (signal) in the 
$m^2$ distributions (see dashed-dot curve in 
Fig.~\ref{Fig1_inv_mass_v2}d):
\begin{equation}
v_2^{(\overline{d})d}(p_T)= \left(v_2^{s+bg}(p_T)-(1-R)v_2^{bg}(p_T)\right)/R,
\label{eq:1}
\end{equation}
where $v_2^{s+bg}(p_T)$ is the measured $v_2$ for $(\overline{d})d$ + 
background at a given $p_T$, $R$ is the ratio signal/(signal+background) 
at that $p_T$, and $v_2^{bg}(p_T)$ is the elliptic flow of the background 
evaluated for $m^2$ values outside of the $(\overline{d})d$ peaks. 

Extraction of the elliptic flow values for the $\phi$ meson 
$(v_{2}^{\phi})$ followed the invariant mass ($m_{inv}$) 
method~\cite{Borghini:2004ra}. For each event, $m_{inv}$, $p_{T}^{pair}$, 
and $\varphi^{pair}$ for each $K^{+}K^{-}$ pair were evaluated. 
Then, for each $p_{T}^{pair}$ bin, 
$v_{2}^{\rm pair} = \left<cos\left(2(\varphi^{pair}-\Phi_{2})\right)\right>$ 
was evaluated as a function of $m_{inv}$ as shown in 
Fig.~\ref{Fig1_inv_mass_v2}c. The value $v_{2}^{\phi}(p_T)$ was then 
obtained from $v_{2}^{\rm pair}(m_{inv})$ via an expression similar to 
Eq.~\ref{eq:1}:
\begin{equation}
\label{v2fit}
	v_{2}^{\rm pair}(m_{inv})=v_{2}^{\phi}
R(m_{inv})+v_{2}^{\rm bg}(m_{inv})(1-R(m_{inv})),
\end{equation}
where $R(m_{inv})=N_{\phi}(m_{inv})/[N_{\phi}(m_{inv})+N_{bg}(m_{inv})]$ 
and $N_{\phi}(m_{inv})$ and $N_{bg}(m_{inv})$ are distributions for the 
$\phi$ meson and the combinatoric background, respectively. 
$N_{\phi}(m_{inv})$ is obtained from the distribution $N_{pair}(m_{inv})$ 
of $K^{+}K^{-}$ pairs from the same event (foreground); $N_{bg}(m_{inv})$ 
is the distribution of pairs obtained from different events with similar 
centrality, vertex, and event plane orientation~\cite{Adler:2004hv}. 
Figure~\ref{Fig1_inv_mass_v2}(a) shows a representative example of the 
latter distributions for $1.6 \le p_{T}^{pair} \le 2.7$ GeV/$c$ and 
reaction centrality 20-60\%. A clear peak signaling the $\phi$ meson is 
apparent in the foreground distribution for $m_{inv} \sim 1.02$ GeV/$c^2$. 
The background distribution was normalized to that for the foreground in 
the region $1.04< m_{inv}< 1.2$ GeV/$c^2$ and subtracted to obtain the 
$N_{\phi}(m_{inv})$ distribution shown in Fig.~\ref{Fig1_inv_mass_v2}(b); 
a relatively narrow $\phi$ meson peak is apparent.

\begin{figure}[htb]
\includegraphics[width=1.0\linewidth]{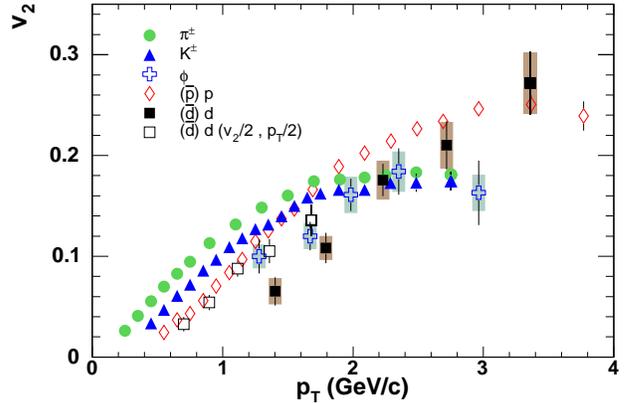}
\caption[]{(color online) 
Comparison of differential $v_2(p_T)$ for $\phi$ 
mesons, $(\overline{d})d$, $\pi^{\pm}$, $K^{\pm}$, and $(\overline{p})p$ 
(as indicated). Results are shown for 20-60\% central Au+Au collisions.
}
\label{Fig2_comp_v2} 
\end{figure}

Determination of the ratio $R(m_{inv})$ was facilitated by fitting this 
distribution with a Breit-Wigner plus a linear function, as shown by the 
solid curve in Fig.~\ref{Fig1_inv_mass_v2}(c). To ensure robust 
$v_{2}^{\phi}$ extraction, the combinatorial background was constructed 
such that $v_{2(mix)}^{\rm bg}(m_{inv})$ gave the same value as 
$v_{2}^{\rm pair}(m_{inv})$ for $m_{inv}$ values not associated with the 
$\phi$ meson peak. Values for $v_{2}^{\phi}$ were extracted via direct 
fits to the $v_{2}^{\rm pair}(m_{inv})$ distribution for each 
$p_{T}^{pair}$ selection (cf. Eq.~\ref{v2fit}). That is, $v_{2}^{\rm 
bg}(m_{inv})$ was parametrized as a linear or quadratic function of 
$m_{inv}$ (depending on the $p_{T}^{pair}$ bin) and $v_{2}^{\phi}$ was 
taken as a fit parameter.

The accuracy of the extraction procedure was verified by checking that the 
$m_{inv}$ dependence of the sine coefficients, $v_{s,2}^{pair}(m_{inv}) = 
\left<sin\left(2(\varphi^{pair}-\Phi_{2})\right)\right>$, were all zero 
within statistical errors. An alternative ``subtraction 
method''~\cite{Pal:2005xy,Adler:2005rg}, in which the raw $\phi$ meson 
yield distribution $dN/d(\varphi_{\phi}-\Phi_2)$ was extracted and fitted 
with the function 
$N(1+2v_2^{\phi}cos\left(2(\varphi_{\phi}-\Phi_2)\right))$, also showed 
good agreement, albeit with larger error bars; $N$ is an arbitrary 
normalization constant.

The differential $v_2(p_T)$ obtained for $(\overline{d})d$ and the $\phi$ 
meson, for centrality 20 - 60\%, are compared to those for $\pi^{\pm}$, 
$K^{\pm}$, and $(\overline{p})p$ in Fig.~\ref{Fig2_comp_v2}. This 
centrality selection was so chosen to (i) maximize the $\phi$ meson signal 
to background ratio over the full range of $p_T$ bins and (ii) enhance the 
distinction between baryon and meson $v_2$ in the intermediate $p_T$ 
range. The shaded bands for $(\overline{d})d$ and the $\phi$ meson 
indicate systematic errors ($\sim 6 - 15\%$), primarily associated with 
the determination of $R$ and $R(m_{inv})$, $v_2^{bg}$ and 
$v_2^{bg}(m_{inv})$ (cf. Eqs.~\ref{eq:1} and \ref{v2fit}), and fitting.

\begin{figure}[tb]
\includegraphics[width=1.0\linewidth]{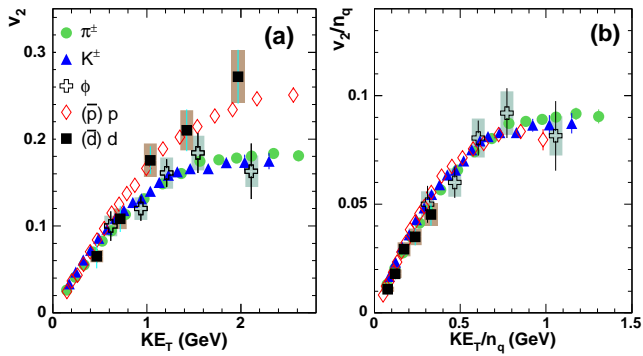}
\caption[]{(color online)(a) $v_2$ vs $KE_T$ for several identified particle 
species obtained in mid-central (20-60\%) Au+Au collisions.  
(b) $v_2/n_q$ vs $KE_T/n_q$ for the same particle species shown in panel (a).
The shaded bands indicate systematic error estimates for $(\overline{d})d$ and 
$\phi$ mesons (see text).
}
\label{Ke_qn_scaling}
\end{figure}

The values for $v_2^{(\overline{d})d}$ shown in Fig.~\ref{Fig2_comp_v2} 
are as much as a factor $\approx 2.5$ lower than those for $\pi^{\pm}$ at 
low $p_T$. This mass ordering pattern 
reflects the detailed 
expansion dynamics of the created matter. As a first test of whether or 
not $v_2$ for $(\overline{d})d$ is additive with respect to its 
constituent hadrons, $v_2^{(\overline{d})d}/2$ vs. $p_T/2$ is compared to 
$v_2^{(\overline{p})p}$ vs. $p_T$. Within errors, they show good agreement 
as would be expected if $v_2^{(\overline{d})d}$ is additive. Another 
salient feature of the results shown is the large magnitude of $v_2$ for 
the $\phi$ meson, which gives an initial indication that significant flow 
development occurs prior to hadronization.

The left and right panels of Fig. \ref{Ke_qn_scaling} compare the unscaled 
and scaled results (respectively) for $v_2$ vs. $KE_T$ for $\pi^{\pm}$, 
$K^{\pm}$, $(\overline{p})p$, $(\overline{d})d$, and the $\phi$ meson, in 
20-60\% central Au+Au collisions. The left panel clearly shows that, 
despite its mass which is comparable to that for the proton, $v_2(KE_T)$ 
for the $\phi$ meson follows the flow pattern of the other lighter mesons 
($\pi$ and $K$), whose cross sections are not OZI suppressed. A similar 
pattern is also observed for the $v_2(KE_T)$ values inferred for 
$D$~mesons (comprised of charmed quarks) from non-photonic electron 
measurements~\cite{Lacey:2006pn,Adare:2006nq}. We interpret these 
observations as an indication that, when elliptic flow develops, the 
constituents of the flowing medium are not ordinary hadrons interacting 
with their standard hadronic cross sections. Instead, they may indicate a 
state in which partonic collectivity dominates the transverse expansion 
dynamics of light, strange, and charmed quarks via a common velocity 
field.

Interestingly, the $v_2(KE_T)$ results shown for the $(\overline{d})d$ and 
the $\phi$ meson are essentially identical at low $KE_T$ ($KE_T \alt 1$ 
GeV), and are in good agreement with those for other charged hadrons, 
including the pion 
with a mass $\sim 13$ times smaller than the 
deuteron. This strengthens the earlier finding that, for low $KE_T$, all 
particle species exhibit the same $v_2$ irrespective of their mass 
\cite{Issah:2006qn,Adare:2006ti,Lacey:2006pn}. The expected difference 
between $(\overline{d})d$ and $(\overline{p})p$ for $KE_T \agt 1$ GeV is 
not tested in Fig. \ref{Ke_qn_scaling}, due to the limited $KE_T$ range of 
the $(\overline{d})d$ data.
	
The right panel of Fig. \ref{Ke_qn_scaling} shows the results for a 
validation test of universal scaling for $v_2(KE_T)$ of baryons and 
mesons~\cite{Issah:2006qn,Adare:2006ti,Lacey:2006pn}. 
The value $n_q = 2\times 3$ is used for 
$(\overline{d})d$ to account for its composite ($p+n$) nature. The scaled 
results shown in Fig. \ref{Ke_qn_scaling}b clearly serve as further 
validation for the experimentally observed universal scaling of $v_2$ for 
baryons and mesons~\cite{Issah:2006qn,Adare:2006ti,Lacey:2006pn}. This 
finding lends strong support to the notion that the high energy density 
matter, created in RHIC collisions, comprise a pre-hadronization state 
that contains the prerequisite quantum numbers of the hadrons to be 
formed. Thus, it appears that partonic collectivity dominates the 
expansion dynamics of these collisions. The special role of $KE_T$ as a 
scaling variable is under investigation.


In summary, we have presented differential $v_2$ measurements for the 
$\phi$ meson and deuteron, and have compared them to those for other 
mesons and baryons. For a broad range of $KE_T$ values, the differential 
$v_2(KE_T)$ for the $\phi$ meson follows the flow pattern for other light 
mesons whose cross sections are not OZI suppressed. The composites 
$(\overline{d})d$ follow the flow pattern for baryons with $v_2$ values 
which are additive. When $v_2/n_q$ is plotted as a function of the 
transverse kinetic energy scaled by the number of valence quarks (ie. 
$KE_T/n_q$), universal scaling results for all particle species measured. 
These observations suggest that the transverse expansion dynamics leading 
to elliptic flow development cannot be understood in terms of ordinary 
hadrons interacting with their standard hadronic cross sections,
but rather in terms of 
a pre-hadronization state in which the flowing medium 
reflects quark degrees of freedom.


We thank the staff of the Collider-Accelerator and 
Physics Departments at BNL for their vital contributions.  
We acknowledge support from 
the Department of Energy and NSF (U.S.A.), 
MEXT and JSPS (Japan), 
CNPq and FAPESP (Brazil), 
NSFC (China), 
IN2P3/CNRS, and CEA (France), 
BMBF, DAAD, and AvH (Germany), 
OTKA (Hungary), 
DAE (India), 
ISF (Israel), 
KRF and KOSEF (Korea), 
MES, RAS, and FAAE (Russia),
VR and KAW (Sweden), 
U.S. CRDF for the FSU, 
US-Hungarian NSF-OTKA-MTA, 
and US-Israel BSF.


\end{document}